# Head-to-Head and Tail-to-Tail Domain Wall in Hafnium Zirconium Oxide: A First Principles Analysis of Domain Wall Formation and Energetics

Tanmoy K. Paul, Atanu K. Saha, Sumeet K. Gupta

*Purdue University, West Lafayette, Indiana, 47907, USA*

Email: paul115@purdue.edu / Phone: (765) 607-3147

180° domains walls (DWs) of Head-to-Head/Tail-to-Tail (H-H/T-T) type in ferroelectric (FE) materials are of immense interest for a comprehensive understanding of the FE attributes as well as harnessing them for new applications. Our first principles calculation suggests that such DW formation in Hafnium Zirconium Oxide (HZO) based FEs depends on the unique attributes of the HZO unit cell, such as polar-spacer segmentation. Cross pattern of the polar and spacer segments in two neighboring domains along the polarization direction (where polar segment of one domain aligns with the spacer segment of another) boosts the stability of such DWs. We further show that low density of oxygen vacancies at the metal-HZO interface and high work function of metal electrodes are conducive for T-T DW formation. On the other hand, high density of oxygen vacancy and low work function of metal electrode favor H-H DW formation. Polarization bound charges at the DW get screened when band bending from depolarization field accumulates holes (electrons) in T-T (H-H) DW. For a comprehensive understanding, we also investigate their FE nature and domain growth mechanism. Our analysis suggests that a minimum thickness criterion of domains has to be satisfied for the stability of H-H/T-T DW and switching of the domains through such DW formation.

## 1. Introduction

Characteristics of domains and their kinetics are crucial for understanding the novel applications of Ferroelectric (FE) material-based devices. The domains in an FE are characterized by a specific direction of electric polarization. In the boundary region between two neighboring domains, defined as the domain wall (DW), polarization transitions from one direction to the other. The nature of the DW and the polarization direction in the domains depend on electrostatic, mechanical and thermodynamic energies and the coupling amongst them. The DWs are commonly indicated by the angle between the polarization of the neighboring domains. Theoretical and experimental works have established a wide variety of DWs in FE materials. For example, with density functional theory calculations and transmission electron microscopy (TEM) experiments, 109°, neutral 180° and 71° DWs in FE $BiFeO_3$ have been demonstrated [1]. 90° DW in tetragonal ferroelectric $PbTiO_3$ has been studied through the combination of first-principles calculations and Landau-Ginzburg-Devonshire (LGD) theory and validated by experiments [2]. 180° charged DW has been experimentally demonstrated in $BaTiO_3$ single crystal through mixed electron/ion screening mechanism [3].

DWs in FEs can be broadly classified into two types: (1) 180° ferroelectric DWs where polarization vectors in two surrounding domains are antiparallel and (2) ferroelastic DWs where polarization vectors are at an angle other than 180°. The 180° FE DWs can be subcategorized into two types. First is the most widely studied 180° neutral DW where DW is parallel to the polarization direction. The second is known as 180° Head-to-Head/Tail-to-Tail (H-H/T-T) DW where DW is in a plane perpendicular to the polarization direction, and polarization vectors of the two domains either face each other (H-H) or away from each other (T-T). DWs can also be classified according to the orientation of polarization within the DW viz. Ising, Neel or Bloch type [4]. Ising type DW is the most common DW in FEs where polarization axis is fixed inside the DW. Neel and Bloch type DWs have also been reported in $Pb(Zr,Ti)O_3$ and $LiTaO_3$ FEs, where rotation of polarization direction happens inside the DW [5] [6].

In 180° H-H/T-T and ferroelastic DW, free carriers may appear at the DW to compensate bound polarization charges. These are known as the charged DWs which have received the attention of scientific community for their prospective applications due to their giant conductivity. This contrasts with the neutral DWs where the conductivity is extremely low. Several theories [7] [8] and experiments [9] [10] have been published on the principle and applications of charged DWs. For example, through electric field-controlled transformation between neutral and charged DW in $BiFeO_3$, ON-OFF switching has been shown experimentally and its application in logic circuits have been proposed [11].

All the examples of different types of DWs mentioned so far are those of conventional perovskite based FEs. Compared to them, doped/undoped Hafnium Oxide ($HfO_2$) is a more recent class of FE material and is widely studied due to its scalability (excellent ferroelectricity in ultra-thin films) and compatibility with complementary metal oxide semiconductor (CMOS) process [12] [13]. Among $HfO_2$ based FEs, 50% Zr-doped $HfO_2$ ($Hf_{0.5}Zr_{0.5}O_2$ or HZO) has been shown to exhibit maximum percentage of FE phase which leads to the highest remnant polarization [14]. Among different types of DW in $HfO_2$-based materials, 180° neutral DW [15] [16] has been by far the most explored type of DW. Besides, non-180° ferroelastic DWs have also been theoretically [17] analyzed and experimentally [18] [19] [20] validated in HZO. On the other hand, stable 180° H-H/T-T DW has been predicted from physical modeling in HZO based devices [21] [22]. For example, previous phase field model for HZO based FE field effect transistor (FEFET) from our research group shows that for a thickness regime greater than 5nm of FE up to 10 nm, stable H-H/T-T DWs can form at suitable voltages during polarization switching via domain nucleation and DW motion [21]. This, in turn, impacts the multi-level memory/synaptic functionality in FEFETs [21].

Despite predicting 180° H-H/T-T DW formation, phase field models do not offer the understanding of microscopic formation, energetics, or domain dynamics of this type of domain wall in HZO. For example, an important aspect of H-H/T-T DW formation is polarization bound charge screening mechanism. Previous theoretical studies have shown that charge compensation of bound charges by free carrier is necessary to form stable H-H/T-T domain wall in FE [7]. But till now, there has been no fundamental investigation from the first principles analysis or experimental demonstration of 180° H-H/T-T DW in HZO.

In this work, we make the following contributions:

➢ We explore the possibilities of 180° H-H/T-T DW formation in HZO-based Metal-FE-Metal (MFM) structures using first principles Density Functional Theory (DFT) calculations.

➢ We show that energy minimization in H-H/T-T DWs occurs through cross-pattern configuration of polar and spacer segments in two neighboring domains of HZO.

➢ We expound the bound polarization charge compensation mechanisms at the interface and DW.

➢ We establish the effect of oxygen vacancy and work function of metal electrode on the stability of H-H/T-T DWs.

➢ We analyze the thickness constraints of H-H/T-T DW formation, their FE polarization profiles and domain growth mechanisms.

## 2. Results & Discussion

We will first discuss the domain configurations, which play an important role in the energetics of the H-H/T-T domain walls. It is widely known that orthorhombic $Pca2_1$ phase (as shown in **Figure 1**) is the source of ferroelectricity in $Hf_xZr_{1-x}O_2$ [23]. The unit cell can be subdivided into two segments along [100] direction: a spacer segment with all the oxygen (O) atoms in the centrosymmetric position and a polar segment with all the O atoms in the non-centrosymmetric position [15]. In different unit cells, the sequence of polar and spacer segments can vary. Also, the atomic arrangement along the directions other than polarization direction can be different [16]. For a given upward polarized unit cell (U), four downward polarized unit cells ($D_a$, $D_b$, $D_c$ and $D_d$) can be derived by performing Euclidean transformations as shown in Figure 1a-1d respectively as explained in our previous work [24]. The performed transformations are as follows: rotation with respect to x-axis ($R_x$), reflection in xy plane ($M_{xy}$) and translation along x-axis ($T_x$). The energy and lattice parameters of all the unit cells in Figure 1 are almost identical. Figure S1 in the Supplementary section shows the switching pathways from upward polarized unit cell to each of these downward polarized unit cells.

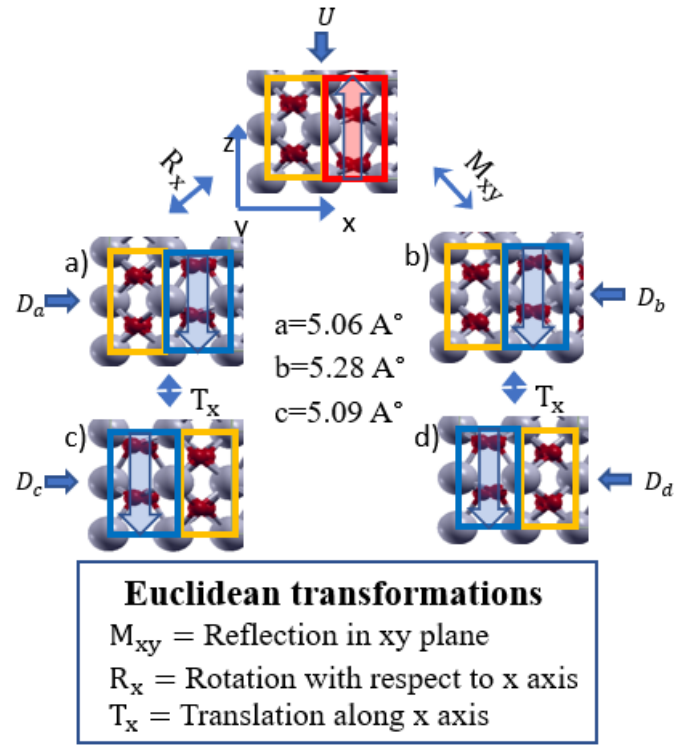

**Figure 1** Orthorhombic $Pca2_1$ phase of $Hf_{0.5}Zr_{0.5}O_2$ with *z*-axis as the polarization direction (Gray and red colored atoms represent Hf/Zr and O respectively). For an upward polarized unit cell (U), 4 downward polarized unit cells ($D_a$, $D_b$, $D_c$ and $D_d$) are derived from Euclidean transformations. Blue and red outlined regions are polar segments with downward and upward polarizations respectively. Yellow outlined regions are spacer segments with centrosymmetric oxygen atoms. The energy and lattice parameters of all the unit cells are almost identical. [24]

From the unit cell atomic configurations in Figure 1, it is evident that the interface with the metal may end with O or Hf/Zr atoms. Moreover, depending on the abundance of O atoms at the interface, different number of oxygen vacancies can form. During the formation of HZO on metal by atomic layer deposition (ALD) through rapid thermal annealing (RTA), O vacancy is created when Hf/Zr atoms are partially oxidized [25]. With proper choice of the electrode and optimization of the fabrication process, oxygen vacancy formation can be minimized [26]. For example, $RuO_2$ electrode has been shown to supply additional oxygen to the HZO film, thus minimizing oxygen vacancies at the interface [27]. To consider the H-H/T-T DW formation in different scenarios, we consider various possible densities of oxygen vacancies at the interface. Note that, due to this assumption, the number of atoms in the HZO sample may correspond to non-integer multiples of HZO unit cell [28].

Let us start with vacancy-free O-ended interface with *Ir* metal and discuss the possibility of T-T DW formation. We will later analyze the effect of oxygen vacancy as well as the possibility of H-H DW formation.

## 2.1 Tail-to-Tail Domain Wall

### 2.1.1 Domain configuration

To understand the attributes of DW formation, we consider all four possible combinations of upward and downward polarized unit cells shown in Figure 1 and form MFM samples with 4.58 nm HZO thickness corresponding to 9 unit cells of HZO. This is the default thickness considered in this work unless otherwise specified. The procedure for forming MFM structures and necessary parameters are described in the Method section. The combination of U with $D_a$, $D_c$ and $D_d$ form MD configurations of $MD_{z,a}$, $MD_{z,c}$ and $MD_{z,d}$ respectively. MD configuration by combining U and $D_b$ is unstable. The SD and stable MD configurations are shown in **Figure 2**a. $MD_{z,a}$ and $MD_{z,c}$ creates local 90° DWs where 90° rotation of Pca2$_1$ phase occurs inside the DW. $MD_{z,d}$ forms *pbcn* phase inside the DW. Atomic configurations inside the DW for all the MD configurations are shown in Supplementary Figure S2. Note that, in $MD_{z,a}$, the sequences of polar and spacer segments are similar in both the domains, and we call it parallel-pattern configuration. On the other hand, in $MD_{z,c}$ and $MD_{z,d}$, the sequence of polar and spacer segments in two domains are alternating in manner. Therefore, we will call them cross-pattern configurations.

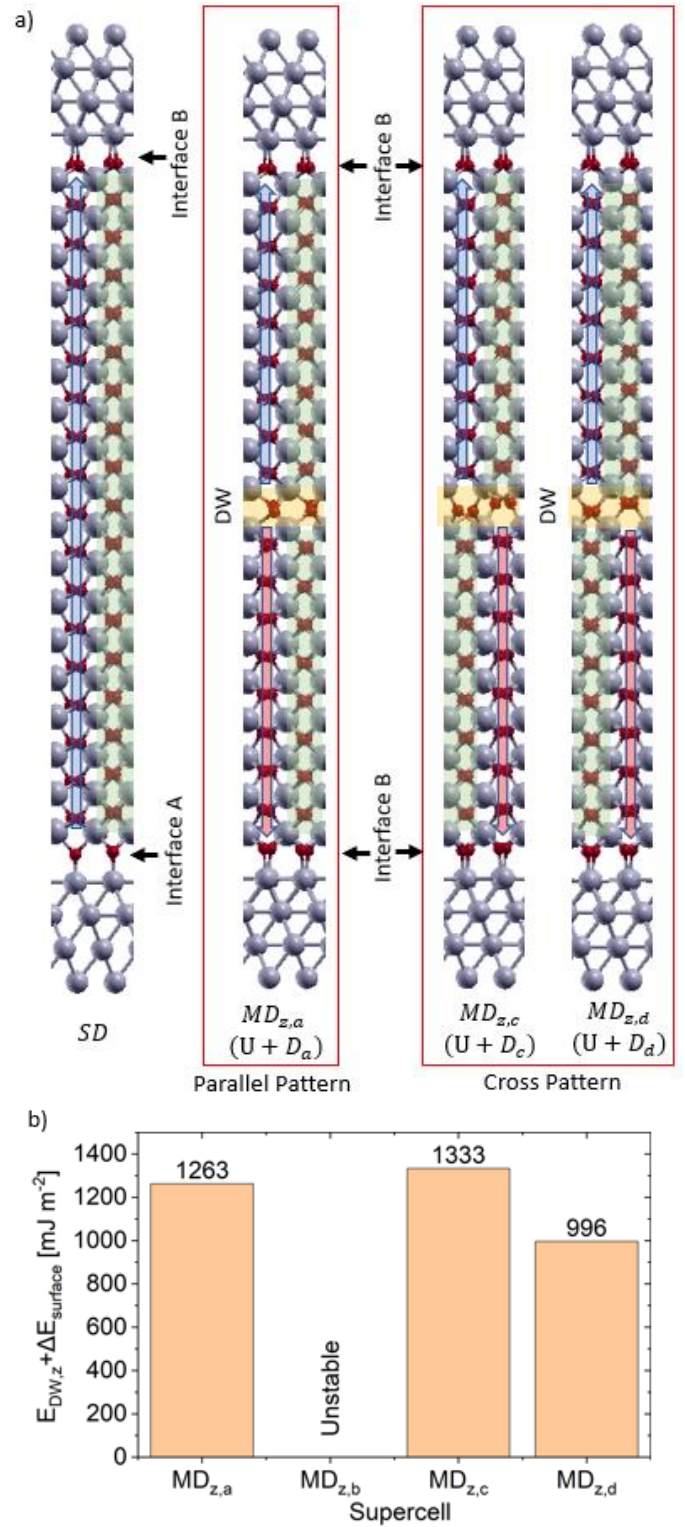

**Figure 2** Atomic configurations of MFM HZO samples for O-ended interface without vacancy a) Single domain and multidomain T-T supercells consisting of U and $D_a$, $D_c$ and $D_d$ of Figure 1 respectively. Combination of U and $D_b$ is unstable. b) Total energy difference with SD for each MD structure. The energy values in the bar include the difference between SD and MD interface energy as well as the DW energy. Thus, cross-pattern $MD_{z,d}$ gives the lowest energy.

The total energy differences per unit cross-sectional area between each of the MD and SD configurations are depicted in Figure 2b. Note that the energy values in the bar include DW energy as well as the difference between MD and SD interface energy. Although what we really want to compare among the MD configurations is DW energy, interface energies cannot be obtained separately. This is because the two interfaces in SD (A and B in Figure 2a) have different polarization bound charges and thus the two interface energies are different in magnitude as described in previous works [29]. In contrast, in each T-T MD configurations, both the interfaces have similar polarization bound charges and similar interfaces. It is also interesting to note that these interfaces are equivalent to the interface B of SD configuration. The similarity of the two interfaces in the T-T MD state suggests that all the three MD configurations have similar overall interface energies. Hence, we expect that the difference between MD and SD interface energy should also be similar for all the MD configurations. Thus, from the energy values in Figure 2b, it is evident that $MD_{z,d}$ has the lowest DW energy. Note that DW here is half *pbcn* unit cell thick (around 0.28 nm). This is much less than the DW thickness in conventional FEs, which typically exhibit H-H/T-T DW width of few nanometers [8] [30]. The thinner nature of such DWs in HZO can be understood intuitively. It is understandable that, near the DW, polarization bound charges exist in the polar segment of HZO, but the spacer segment is free of polarization bound charges. In the cross-patterned $MD_{z,d}$ configuration, polar segment faces spacer segment of the opposite domain near the DW. Thus, similar bound charges near the DW avoid facing each other and minimize the energy cost. Consequently, even half unit cell thick DW can have sustainable electric fields. Although $MD_{z,c}$ also exhibits cross-pattern configuration, the strain related to 90° rotation of Pca2$_1$ phase on both sides of the DW makes it energetically costlier than $MD_{z,d}$. Henceforth, we will focus on the cross-pattern configuration of $MD_{z,d}$ as the minimum energy T-T DW. With this understanding of the energetics, we will now delve deeper into its bound polarization charge screening mechanism.

### 2.1.2 Bound polarization charge screening

In T-T configuration, negative and positive bound polarization charges appear at the DW and interface respectively. To obtain stable T-T DW in HZO, it is necessary to have proper screening of these charges.

**2.1.2.1 Screening at interface.** It is well known that metals have excellent capability to compensate bound charges in ferroelectric interfaces through their mobile charges [31] [32]. This is also true for the MFM structures discussed in this work. To validate this claim, we will discuss the atomic layer-wise Projected Density of States (PDOS) of HZO SD configuration with *Ir* metal. **Figure 3**a shows PDOS for the two atomic layers of HZO near the bottom (A) and top (B) interfaces. Here, layer-wise conduction band (CB) and valence band (VB) shift is negligible between the two interfaces. Thus, depolarization field is negligible within the domains. This suggests that bound polarization charges at the two interfaces of SD are screened by the metal very well. As discussed in section 2.2.1, both the interfaces in the T-T configuration are similar to one of the interfaces of SD. Thus, it can be inferred that metals compensate interface bound charges in T-T configuration as well. Note that in SD configuration, VB maximum and CB minimum are sufficiently away from the metal Fermi level. This reiterates that HZO acts as an excellent insulator in the SD configuration.

**2.1.2.2 Screening at DW.** Now, we will investigate the bound charge screening mechanism near the DW. To screen negative bound polarization charges at the T-T DW, positive mobile charge carriers (holes) should appear. This can be observed from the total PDOS inside FE HZO and at the central oxygen layer shown in Figure 3b and 3c, respectively. Here, the maximum of the VB comes closer to the Fermi level to generate

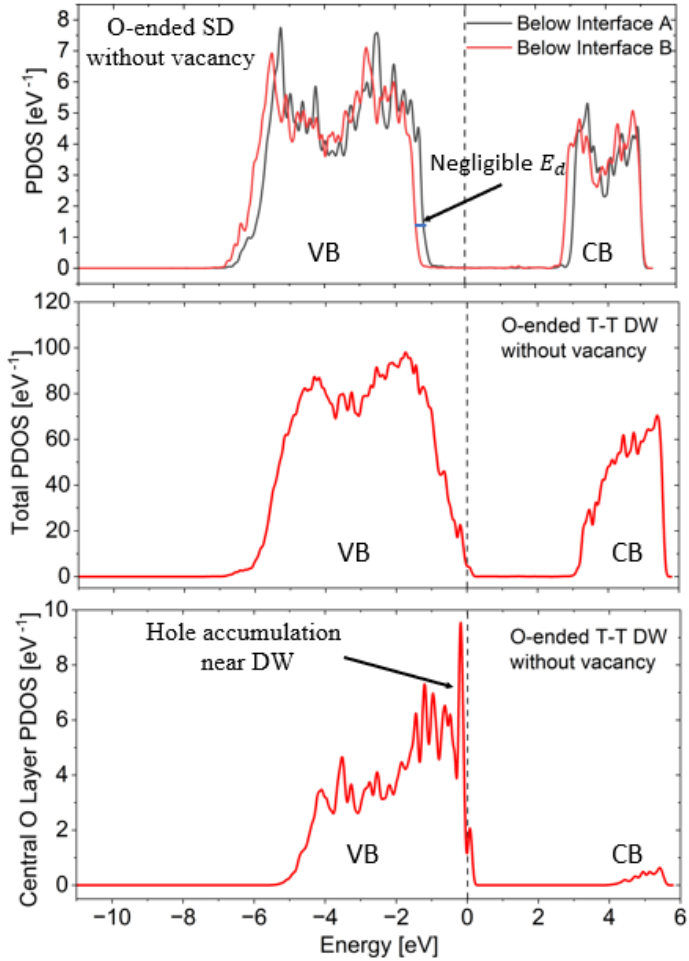

**Figure 3** a) Projected Density of States (PDOS) of atomic layers of SD HZO near the interfaces A and B in Figure 2a. The small shift of CB and VB between the two atomic layers suggests negligible depolarization field i.e., excellent bound polarization charge screening near the interfaces by metal. b) Total PDOS and c) PDOS at central oxygen layer of MD T-T configuration in HZO. The VB maximum and fermi level suggest that holes appear at the DW to screen negative polarization bound charges. This comes at the cost of band bending, i.e., depolarization field.

holes. The hole density is maximum in the O layer at the center of the DW. PDOS at each layer of HZO sample is shown in Supplementary Figure S3. The layerwise VB shift in Figure S3 indicates that the accumulation of holes happens by band bending inside the FE layer. Depolarization field acts as the source of this band bending. But excessive depolarization field destabilizes FE phase. Thus, for T-T DW to form, an optimized amount of band bending is needed inside the FE such that (a) sufficient mobile charge carrier density can be generated to compensate the bound charges in the T-T DW and (b) depolarization field is weak enough so as not to destabilize the DW.

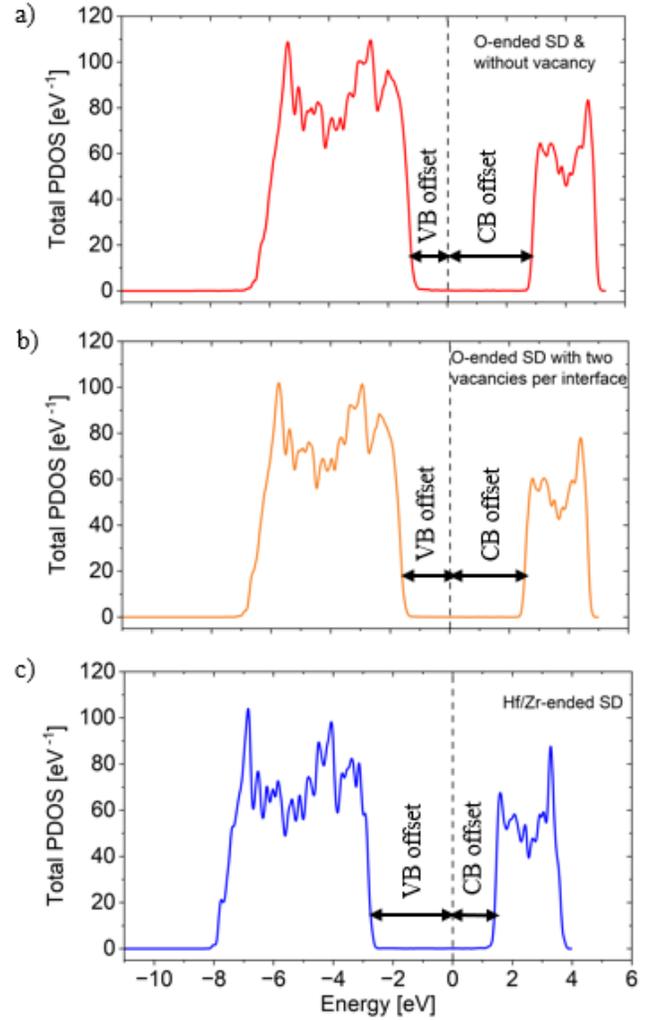

**Figure 4** Total PDOS of atomic layers of SD HZO for a) O ended interface without oxygen vacancy b) O-ended interface with two oxygen vacancies per interface c) Hf/Zr ended interface. Oxygen vacancy increases VB offset and decreases CB offset.

### 2.1.3 Effect of oxygen vacancy

So far, we have discussed the stability of T-T DW for a vacancy-free O-ended interface. Now we will discuss the effect of oxygen vacancy on its stability. For that, let us first understand its effect on SD configuration.

**2.1.3.1 Band offset modulation due to oxygen vacancy.** The presence of oxygen vacancy increases the electro-positivity of HZO [33]. Thus, it decreases the likelihood of removing an electron and increases the likelihood of adding an electron into the HZO sample. This means that both the electron affinity and ionization potential increase. As a result, oxygen vacancy shifts the VB and CB down, i.e., VB offset with respect to the metal electrode increases and CB offset decreases. The band offsets are important factors for stabilization of H-

H/T-T DWs as will be evident in subsequent sections. To understand the O-vacancy effect on band offsets, we perform the SD band calculation for three different types of HZO termination: O-ended interface with no oxygen vacancy, 2 oxygen vacancies per interface and Hf/Zr-ended interface as shown in Supplementary Figure S4. The total PDOS of these FE samples are shown in **Figure 4**a-c respectively. For oxygen vacancy free interface with *Ir* metal, VB offset is smaller compared to CB offset as shown in Figure 4a. For Hf/Zr-ended interface in Figure 4c, CB offset becomes smaller compared to VB offset. The PDOS of HZO layers near the two interfaces for each of the SD configurations are shown in Supplementary Figure S5. The plots suggest negligible depolarization field inside the FE in each of the SD configurations. Thus, polarization bound charges at the interfaces are well-screened by the metal electrode, irrespective of the interface termination. This means that charge screening at the DW is the most crucial factor for the stability of T-T DW.

**2.1.3.2 Stability of T-T DW with oxygen vacancy.** With this understanding of the band offset modulation with O vacancy, let us now analyze its effect on the stability of T-T DW. For that purpose, we first take a stack of metal-FE HZO. Then, we relax the interface by fixing the furthest HZO unit cell from the interface to the bulk HZO atomic configuration. The interface relaxation ensures that bound polarization charges near the interface get compensated. We perform this relaxation for different stacking order (metal on top or bottom), number of oxygen vacancies (zero to four), oxygen vacancy distribution (vacancy in polar or spacer layer) and polarization direction (up and down) of HZO. Then we clamp two stacks to form necessary MFM structures which act as initial atomic configurations of T-T DW with oxygen vacancy. The metal-HZO stacks and clamping plane is shown in Supplementary Figure S6a.

To understand the stability of these unrelaxed DWs, we perform PDOS calculations. **Figure 5**a shows the PDOS of the atomic layers near the interface and at the DW considering interfaces without oxygen vacancy. As described in section 2.1.2.2, holes appear at the T-T DW to screen bound charges at the cost of the depolarization field. The magnitude of the depolarization field is determined by the shift of the VB maximum as shown by the arrow in Figure 5a. Now, as discussed in section 2.1.3.1, when there is an increased number of oxygen vacancies at the interface, the VB offset increases. Thus, it requires larger depolarization field to accumulate holes at the DW. This increase in the depolarization field is visible from the larger arrow size in Figure 5b compared to that in Figure 5a. If the domains can withstand this increased depolarization field, the T-T DW becomes stable. Otherwise, they transform to single domain. Our calculation for the 4.58 nm thick HZO suggests that if a single vacancy per interface forms at the spacer layer with *Ir* metal, it relaxes to a stable T-T DW. Further, as we increase the number of oxygen vacancies per interface, T-T DW becomes unstable. On the other hand, for vacancies at the polar layer, the T-T DW becomes unstable after relaxation even with a single vacancy.

To understand the attribute of this spatial distribution of oxygen vacancy, we compare the PDOS of atomic layers below the interface for O vacancies in both polar and spacer layer (Figure 5c). It suggests that VB offset is slightly larger when vacancy is in the polar layer as opposed to when vacancy is in the spacer layer. As a result, depolarization field needed for hole accumulation at the DW is also higher when O-vacancy is in the polar layer. Thus, although T-T DW forms when a single O-vacancy exists in the spacer layer, it destabilizes when the O-vacancy is in the polar layer. It is important to point out that the difference in the VB offsets between polar and spacer layers in the unrelaxed condition is minor and further investigation is needed to analyze other possible causes.

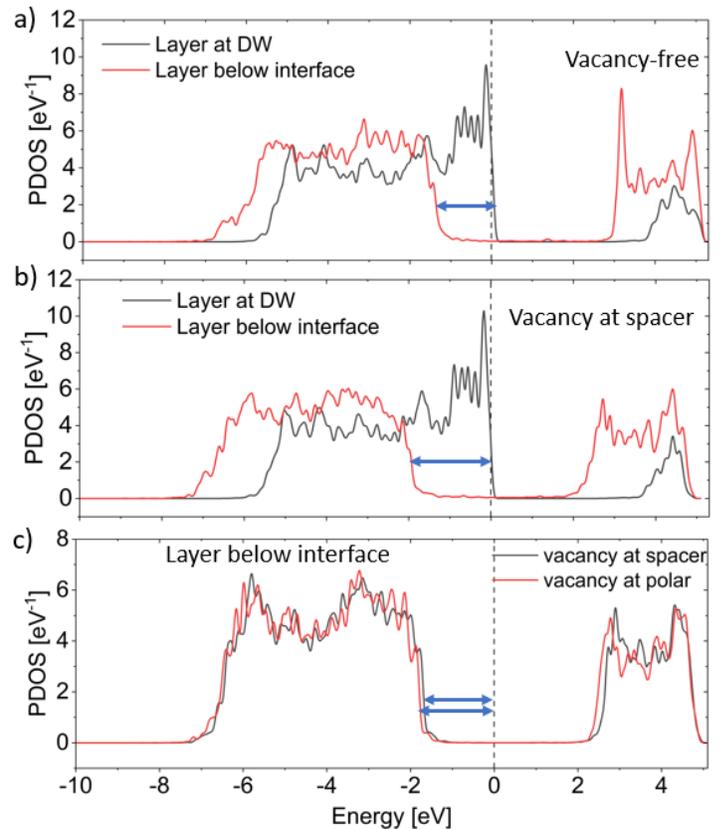

**Figure 5** PDOS of atomic layers near the interface and at DW of unrelaxed T-T configuration with relaxed interfaces for a) O-ended interface without vacancy and b) O-ended interface with a vacancy at the spacer layer. c) Comparison of PDOS near the interface between vacancy at polar and spacer layer. Depolarization field needed for creating holes at the DW is more when there is a vacancy at the spacer layer compared to vacancy-free interface. The field is slightly larger when the vacancy is at the polar layer.

### 2.1.4 Effect of metal electrode

Besides the oxygen vacancy, stability of T-T DW also depends on the work function of the metal electrode. The higher work function of metal shifts the fermi level down with respect to the vacuum level and thus reduces the VB offset of HZO. Then, the depolarization field for accumulating holes reduces and the stability of T-T DW increases. On the other hand, for low work function metal, VB offset increases and it reduces the T-T DW stability. We verified this dependence on metal work function by using tungsten (W) as low work function metal and Platinum (Pt) as high work functional metal. With Pt electrode, stability of T-T DW with respect to vacancy is like what was observed in *Ir* metal because of their similar work functions. With W electrode, T-T DW is unstable even with O-vacancy-free interface. The results suggest that in FE devices using high work function metals, locally existing oxygen vacancy free interface or interface with single oxygen vacancy at the spacer layer act as the nucleation site for T-T DW formation.

### 2.2 Head-to-Head Domain Wall

Now, we will briefly discuss the stability of H-H DW formation. Let us start with H-H configurations for vacancy-free O-ended interface with *Ir* metal. Following the process discussed in the section 2.1.3.2, we start with relaxed interfaces of metal-HZO with bulk HZO atomic coordinates away from the interface and clamp two stacks to form initial H-H configurations (Supplementary Figure S6b). Then we calculate PDOS for these unrelaxed structures. As opposed to hole accumulation in T-T DW, for H-H DW, electrons accumulate at the DW to screen positive bound polarization charges at the cost of depolarization field. Here, shift of the CB represents the required depolarization field (as opposed to the shift of VB in case of T-T DW). Recall from section 2.1.3.1 that for vacancy-free O-ended interface with *Ir*, CB offset is much larger than VB offset. Thus, to accumulate free carriers near the DW, larger CB shift is needed compared to VB shift. As a result, larger depolarization field is needed in H-H DW compared to T-T DW to screen bound charges at the DW. The PDOS of atomic layers below the interface and near the DW for both configurations are shown in **Figure 6**. The width of the arrows in Figure 6a (T-T DW) and 6b (H-H DW) represent the corresponding depolarization field within the domains. With such higher depolarization field of H-H DW, it becomes unstable after relaxation [34] [35].

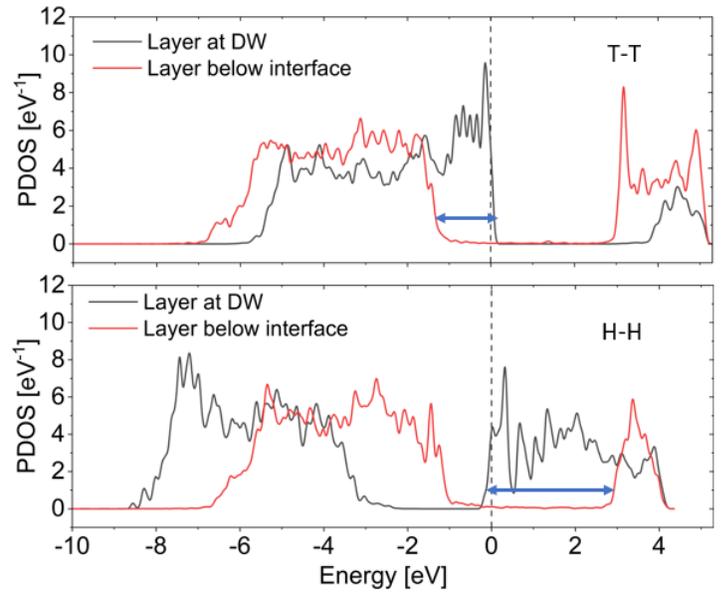

**Figure 6** PDOS of atomic layers below the interface and at the DW of unrelaxed a) T-T and b) H-H configurations with relaxed vacancy-free O-ended interface. The arrows represent the bend bending needed to accumulate necessary free carriers for bound charge compensation at the DW. Thus, higher depolarization field is needed for stabilizing H-H DW.

As we increase the number of O vacancies at the interface, CB offset reduces. Thus, depolarization field needed for screening bound charges at the H-H DW reduces. This increases the stability of H-H configuration. It turns out that for higher work function metals (i.e., *Ir, Pt)*, H-H DW becomes stable only when the interfaces are Hf/Zr-ended. For lower work function metal, CB offset reduces and it increases the H-H DW stability. For W electrode, H-H is stable for all interfaces including the O-ended interface without O vacancy. The relaxed H-H DW configuration with Hf/Zr-ended interface for *Ir* and corresponding PDOS is shown in Supplementary Figure S7.

From the discussions made so far, it is evident that both H-H and T-T configurations in HZO carry mobile carrier (electron and hole respectively) near the DW region. (This may be similar to the charged DW in perovskite FEs [9] and calls for further investigation.) In addition, reduction of oxygen vacancy at the interface and using metal electrode of high work function is favorable for hole accumulation at the DW and stabilization of T-T DWs. On the other hand, increased oxygen deficiency at the interface and low work function metal favors H-H DW formation through electron accumulation at the DW. Polarization switching can occur through either H-H or T-T configuration depending on the density of oxygen vacancy at the interface and electrode selection. Previous research works have predicted O enriched interfaces to be more energetically favorable than Hf/Zr ended interface [29]. Thus, for $Ir$, H-H DW is less likely to form compared to T-T DW. As we are focusing on $Ir$ metal in this study, for the rest of the paper, we will focus on T-T DW and analyze the characteristics of T-T DW with vacancy-free O-ended interface.

### 2.3 T-T DW Characteristics

#### 2.3.1 DW stability with HZO thickness

As discussed earlier, stability of T-T configurations depends on the magnitude of depolarization field. To obtain the magnitude of this electric field, we calculate plane averaged local and macroscopic electrostatic potential (EP). EP for our 4.58 nm thick sample is shown in **Figure 7**. Figure 7a shows macroscopic EP in the SD state which is almost flat. Thus, the depolarizing electric field inside the domain of the SD state is negligible as discussed before with PDOS. In the MD state, the EP is shown in Figure 7b when the domains are of equal width, i.e., the DW is at the center. As discussed earlier with PDOS, band bending occurs in MD due to the depolarization field. The depolarization field is indicated by the non-zero slope of the macroscopic EP inside the domains. The depolarization field is the same in both the domains because of the equal width of the domains and similar interfaces on both sides.

Now, the depolarization field changes as the DW moves away from the center. Structural relaxation reveals that DW can shift by half-unit cell in T-T domain configuration, i.e., domains can consist of even or odd number of half-unit cells along the polarization direction. The EP with DW shifted by two half-unit cells (one unit cell) from the center is shown in Figure 7c. As the DW shifts from the center, the slope of the

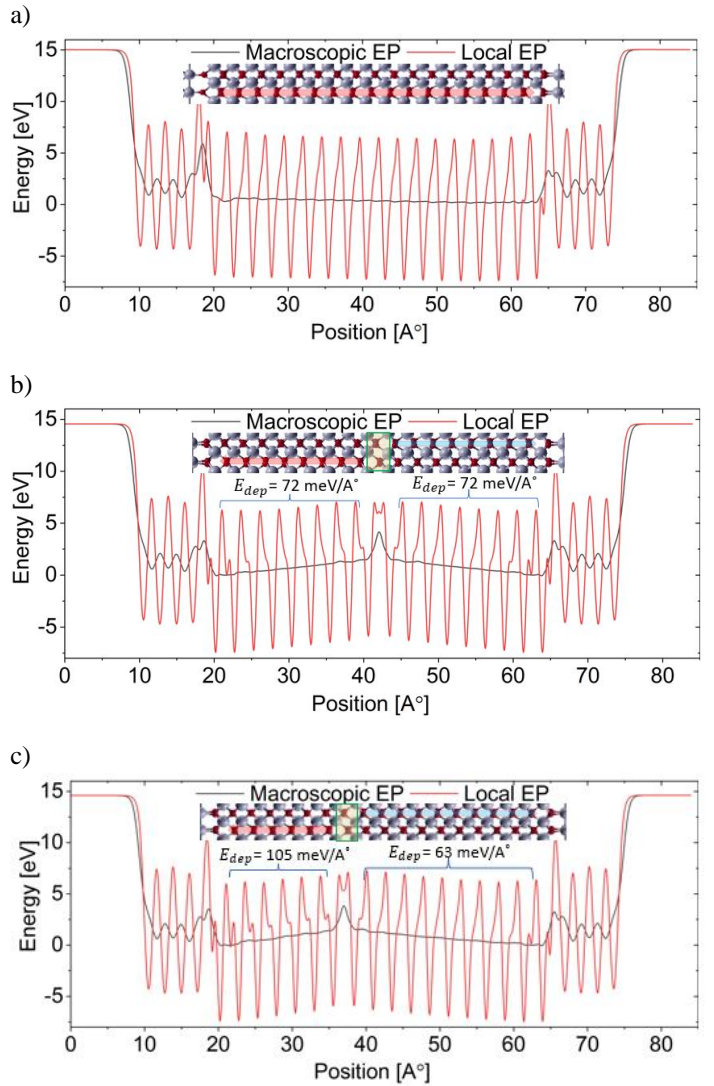

**Figure 7** Plane averaged local and macroscopic electrostatic potential along the polarization direction for 4.58 nm thick HZO in MFM when a) HZO is in SD state b) DW is at the center c) DW is one unit cell away from the center. Insets show the corresponding atomic configurations. At SD state there is negligible depolarization field as opposed to finite values in MD state. Depolarization field increases in the narrower domain when DW shifts from the center.

macroscopic EP on the narrower domain increases, indicating a larger depolarization field. The depolarization field in the narrower domain for different locations of DW is shown in **Figure 8**a. The values along the horizontal axis are the locations in terms of half-unit cell with respect to the center. When the depolarization field in a domain exceeds a critical value ($E_{cr}$), the MD configuration becomes unstable. Hence, it transforms to the SD state. This critical electric field sustained by the domains in T-T configuration ($E_{cr}$) is obtained as 0.105 V/A° (10.5 MV/cm) as seen in Figure 8a. This is the case when thickness of the narrower

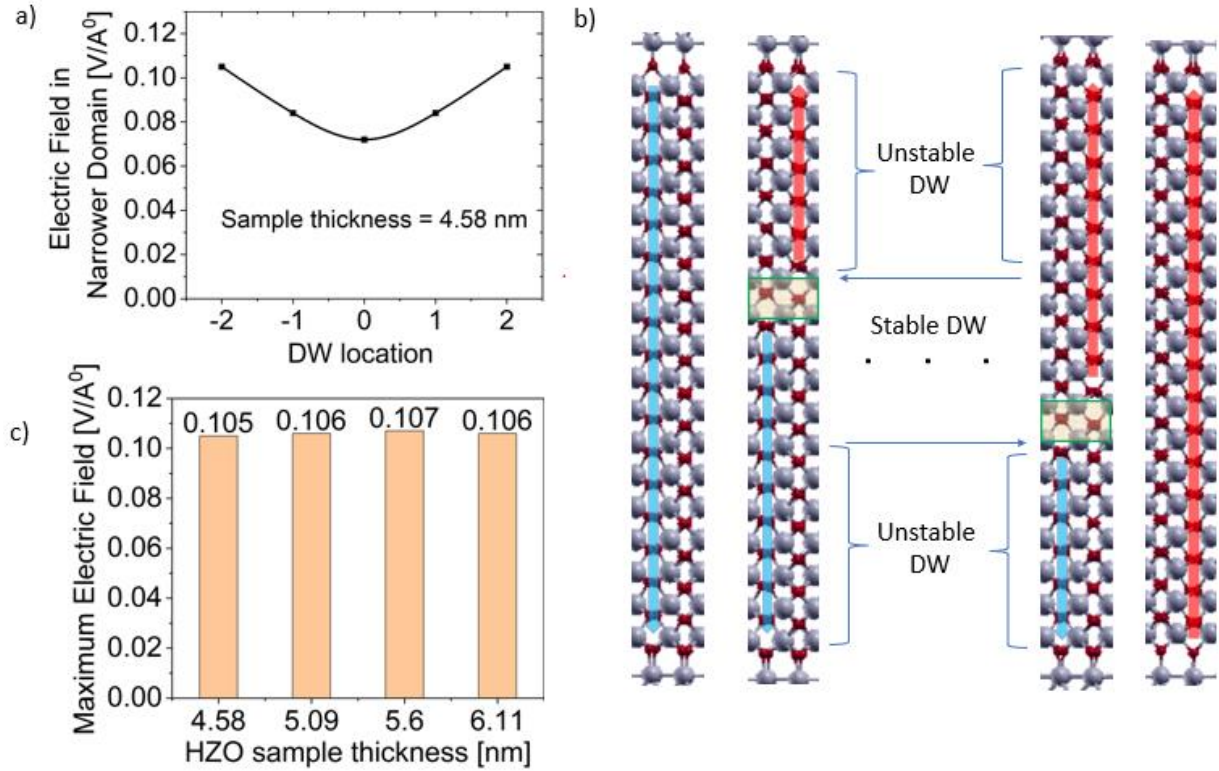

**Figure 8** a) Depolarization field in the narrower domain with the position of the DW away from the center in terms of half-unit cell. b) Regime of stability of DW. DW is only stable within a certain distance from the center. Beyond that point, the excess depolarization filed destabilizes the DWs. c) Consistency of the magnitude of maximum depolarization field with varying HZO sample thickness.

domain reaches its critical thickness ($t_{cr}$) of 1.63 nm (thickness of 3 unit cells). Thus, T-T DW exists when it is located at least 3 unit cells (~1.63nm) away from both the interfaces. Hence, HZO sample thickness needs to be more than 3.13 nm (thickness of 6 unit cells) for observing T-T DWs (given that other conditions, as discussed before, are met). The regime of stability of the T-T DW is depicted in Figure 8b. We take HZO samples from 4.58 nm to 6.11 nm (thickness of 9 unit cells to 12 unit cells) and calculate the critical field, $E_{cr}$ for each of the samples. $E_{cr}$ of 10.5 MV/cm and $t_{cr}$ of 1.63 nm is obtained irrespective of the sample thicknesses as shown in Figure 8c. Thus, the critical domain thickness condition for T-T DWs persists for different sample thicknesses of HZO. Note that $t_{cr}$ with single oxygen vacancy at the spacer layer is slightly higher and is around 2.15 nm. We also observe that it is possible to obtain stable T-T DWs, with a single O-vacancy in the polar layer if the HZO thickness is increased. The value of $t_{cr}$ is around 3.17 nm for T-T DW considering a single O vacancy at the polar layer.

### 2.3.2 Polarization profile

To show how polarization evolves inside the domains for the T-T configuration, we plot the polarization profiles in Figure 9 for vacancy-free O-ended interface. Note that polarization of the O layer at the interface is not calculated. **Figure 9**a shows the polarization profile in the SD state. Here, due to negligible depolarization field (Figure 7a), polarization is almost equal to bulk polarization (49 µC cm$^{-2}$) throughout the domain. On the other hand, in the MD state, i.e., when T-T DW exists, polarization is suppressed in each domain due to the depolarization field. When the DW is at the center and depolarization field is 7.2 MV/cm (from Figure 7b), polarization in the domains is 35 µC cm$^{-2}$. When the DW shifts from the center, depolarization field increases in the narrower domain and decreases in the wider domain as stated earlier. As a result, local polarization decreases in the narrower domain and increases in the wider domain. At depolarization field of 10.5 MV/cm (from figure 7c), polarization reduces to 25 µC cm$^{-2}$ in the narrower domain. The polarization profiles for DW at the center and one unit cell away from the center are shown in Figure 9b and 9c respectively.

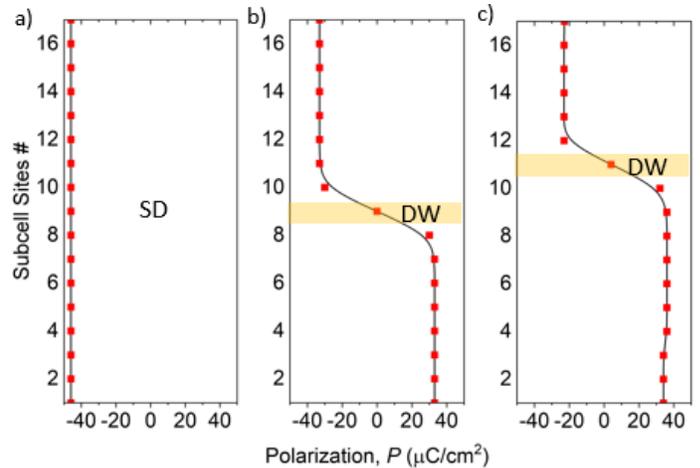

**Figure 9** Atomic layer wise Polarization profile along the polarization direction when a) the sample is in SD state b) DW is at the center c) DW is one unit cell away from the center. Polarization is close to bulk value in single domain but decreases with increasing depolarization field in the multidomain states. Interface oxygen layer has been discarded for polarization calculation.

### 2.3.3 Domain nucleation and DW motion

With the understanding of the regime of stability of T-T DW discussed in section 2.3.1, let us now look at the domain switching pathway through T-T DW formation. As described earlier, minimum distance of the DW from an interface with *Ir* metal electrode is 1.63 nm. This means that the nucleation site must exceed the critical thickness of 1.63 nm (3 unit cells). After nucleation, DW motion occurs by a quantum of half-unit cell along the polarization direction. This continues until DW reaches the critical thickness from the other side, after which it collapses to the SD state. The switching process is depicted in **Figure 10**a. The corresponding energy barrier profile for this switching pathway obtained from NEB calculation is shown in Figure 10b. As can be seen, the nucleation barrier height is 1.67 eV which corresponds to the reversal of 3 unit cells. Thus, the nucleation barrier per unit cell reversal is approximately 0.55 eV (1.67/3 eV). The DW motion barrier for half-unit cell reversal is 0.12 eV (0.12×2 = 0.24 eV per unit cell). Previous works have shown nucleation and DW motion barrier height of 1.34 eV per unit cell for the widely studied lateral DWs along *x*-axis [15]. Thus, the nucleation and DW motion energy barrier per unit cell along polarization direction for T-T DWs is significantly less than that for lateral DWs along the *x* direction. According to Merz's law, DW speed is proportional to $\exp(-\frac{E_a}{E})$, where $E_a$ is the activation field corresponding to barrier height and $E$ is the applied electric field. This means that domain nucleation time is expected to be much lower and DW motion to be much

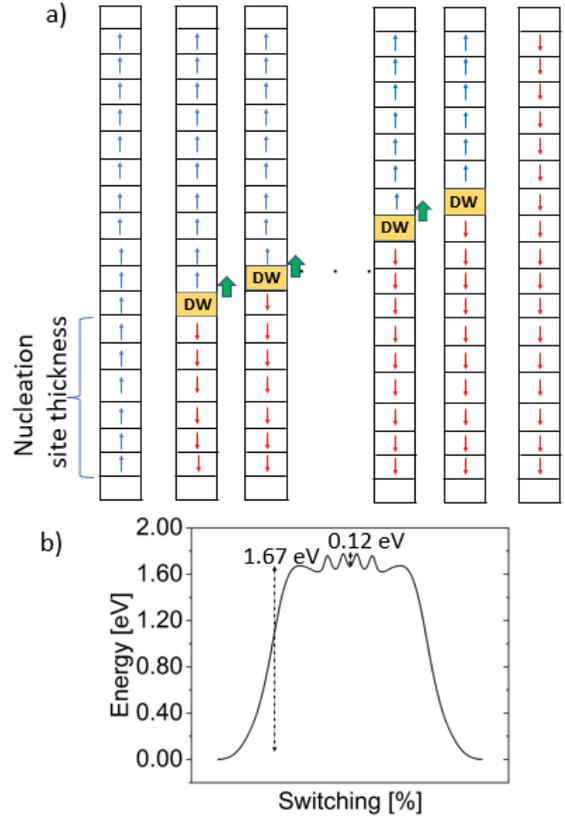

**Figure 10** a) Schematics of nucleation and DW motion for polarization reversal through T-T DW. Arrow in each slot represents polarization of half-unit cell along the polarization direction. The empty slot represents interface oxygen layer. Minimum nucleation site thickness is 1.63 nm (3 unit cell thick). b) Barrier height for polarization reversal. Nucleation barrier per unit cell is 1.67/3 = 0.55 eV and DW motion barrier per unit cell is 0.12×2 = 0.24 eV.

faster along the polarization direction compared to the lateral (*x*) direction.

In summary, using first principles density functional theory, we analyzed T-T and H-H DW formation in HZO based Metal-Ferroelectric-Metal structures. We found that stable T-T configuration with *pbcn* like phase near the DW can form for cross patterned polar and spacer segments in the neighboring domains. Low density of oxygen vacancy and higher work function of metal electrode enhances the favorability of T-T DW formation. On the other hand, stable H-H configuration is more likely to form with high density of oxygen vacancy or low work function of metal. Bound charges near the DW are screened by mobile carriers originating from band bending due to depolarization field. A balance between mobile carrier accumulation and depolarization field is needed for the stability of H-H/T-T DWs. Our detailed analysis with T-T configuration shows that DW formation is only possible above a critical thickness (3.13 nm for Ir-HZO-Ir) of the HZO sample. This is attributed to the increase in the depolarization field for low thicknesses. Based on the polarization reversal barriers, domain growth is expected to be faster along the polarization direction compared to those of the widely studied lateral direction. The unique results obtained in the work will help the ferroelectric research community to understand the possibility of H-H/T-T DW formation in HZO. This will be critical for analyzing the polarization switching process and may lead to the applications of H-H/T-T DW in HZO.

**Methodology**

We use Quantum Espresso software package [36] for First Principles Density Functional Theory calculations and Xcrysden software package [37] for presenting atomic configurations in this work. In Quantum espresso, Projector-Augmented Wave (PAW) pseudopotentials with non-linear core correction and Perdew–Burke–Ernzerhof generalized gradient approximation (GGA-PBE) as exchange correlation functional is used. $HfO_2$ unit cell of ferroelectric orthorhombic $Pca2_1$ phase consists of 4 Hf atoms and 8 oxygen atoms as shown in Figure 1. Among them, non-centrosymmetric O atoms are responsible for the induced dipole moment. In HZO, half of the Hf atoms are replaced by Zr dopant. The orthorhombic $Pca2_1$ unit cell of HZO is optimized until $10^{-6}$ Rydberg (Ry) error in ionic minimization energy and $10^{-4}$ Ry Bohr$^{-1}$ force tolerance of all directional components are reached. Kinetic energy cut-off of 60 Ry for wavefunction and 360 Ry for charge density and potential are used with proper convergence test of the parameters. Brillouin zone is sampled with $6 \times 6 \times 6$ Monkhorst-Pack grid of k-points. The calculated relaxed lattice parameters of the unit cell (5.06 A°, 5.28 A° and 5.09 A° along the three orthogonal directions) match with previous theoretical calculations and experiments [23] [14]. We obtain the FE polarization from ionic and electronic contributions calculated using berry phase approach of Kohn Sham states (as per the modern theory of polarization) [38]. We obtain spontaneous polarization of 50 μC cm$^{-2}$ with polarization quantum of 120.24 μC cm$^{-2}$ along the z-direction consistent with previous works for $HfO_2$ [17]. It should be mentioned that experimentally observed residual polarization (21 μC cm$^{-2}$ after wake-up [39]) is lower due to the polycrystalline nature of HZO in experimental samples [40]. Other than $Pca2_1$ phase, $Fm\overline{3}m$, $pbcn$ and $P4_2/nmc$ phases also appear at the DW or during polarization switching. For comparison of atomic configurations with these phases, we obtain their unit cells and associated FE polarization using similar criteria as $Pca2_1$ phase.

For previously reported lateral HZO DWs, periodically repeated supercells are taken along all directions [24]. However, in this work, for H-H/T-T DW, we use vacuum on both sides along the z-axis (polarization direction) to form Metal-Ferroelectric-Metal structures. Unless otherwise specified, we use Iridium (*Ir*) as the metal due to its applicability in HfO2-based ferroelectrics and its low lattice mismatch with HfO2. Cubic $Fm\overline{3}m$ unit cell of *Ir* is first relaxed to obtain the optimized configuration. The bulk *Ir* is then cleaved along the plane represented by miller indices (111). A (2×2) supercell of the cleaved surface has in-plane lattice parameters of 5.48 A° and 4.75 ° where all the angles are 90°. This *Ir* supercell is then strained with 3.6% compressive strain and 6.3% tensile strain respectively along the two in-plane directions to match with the underlying (1×1) sized HZO. This reasonably low lattice mismatch allows us to simulate small supercells as well as obtain realistic bonding scenario near the HZO-*Ir* interface. Convergence test with respect to the metal thickness on both sides is performed. It is observed that metal atomic configurations are affected within a very thin region of the interface for different metal thicknesses. After convergence test, we take 7 A° of metal on each side. Then, HZO is sandwiched between metal on both sides. Due to the inherent periodic nature of structures in the DFT methodology, the vacuum around the model should be wide enough to avoid interactions between two contiguous (but unintended) structures. Vacuum of 10 A° on both sides of the MFM structure is formed giving a total of 20 A° gap. The validity of the assumed vacuum thickness is verified by energy convergence with respect to vacuum layer thickness. Energy error of $10^{-5}$ Ry, force tolerance of $10^{-4}$ Ry Bohr$^{-1}$, $2 \times 2 \times 1$ Monkhorst-Pack grid points and Gaussian smearing of 0.01 Ry are used for calculation of MFM. This low k-point density increases the computational efficiency significantly with some sacrifice in the accuracy of the system energies. However, we verify that energy differences - the quantities that we are mostly interested in – are unaffected by these assumptions. Single domain slabs and multi-domain slabs of H/H-T-T configurations are ionically relaxed by constraining the in-plane lattice parameters to be equal to bulk single domain lattice parameters. On the other hand, the out-of-plane lattice parameter gets relaxed due to the vacuum on both sides. Despite the wide vacuum, there might be unintended spurious electric field from the neighboring out-of-plane supercells. To nullify its effect, we apply dipole correction through a fictitious dipole when needed.

We use Nudged Elastic Band (NEB) method for finding energy barrier profiles of nucleation and DW motion. The Projected Density of States (PDOS) and local 3-dimensional Electrostatic Potential (EP) are obtained from data postprocessing packages in Quantum Espresso. The atom wise PDOS data obtained from Quantum Espresso are post-processed in MATLAB to obtain layer wise PDOS. Due to the widely known bandgap underestimation issue with GGA functional, the band offset values from PDOS are not exact in magnitude but they provide a good understanding of the trends. To get more accurate values of depolarization field, EP is used in MATLAB to extract the plane-averaged local electrostatic potential. For macroscopic potential, an averaging window size equal to the interlayer oxygen spacing is applied to minimize oscillations. The magnitude of electric field in a region is defined by the slope in the macroscopic potential.

# References

[1] "Y. Wang, et al, " BiFeO3 Domain Wall Energies and Structures: A Combined Experimental and Density Functional Theory +U Study", Phys. Rev. Lett. 110, 267601, 2013".

[2] "Y. J. Wang, Y. L. Tang, Y. L. Zhu, Y. P. Feng, X. L. Ma, "Converse flexoelectricity around ferroelectric domain walls",Acta Materialia, Volume 198, 1 October 2020, Pages 257".

[3] "P. Bednyakov, T. Sluka, A. Tagantsev, et al., "Formation of charged ferroelectric domain walls with controlled periodicity", Sci Rep 5, 15819 (2015)".

[4] "D. M. Evans, V. Garcia, D. Meier, M. Bibes, "Domains and domain walls in multiferroics", Phys. Sci. Rev., vol. 5, no. 9, 2020, pp. 20190067".

[5] "X.-K. Wei, C. L. Jia, T. Sluka, et al., "Néel-like domain walls in ferroelectric Pb(Zr,Ti)O3 single crystals", Nat Commun 7, 12385 (2016)".

[6] "S. C.-Hertel, H. Bulou, R. Hertel, et al., "Non-Ising and chiral ferroelectric domain walls revealed by nonlinear optical microscopy", Nat Commun 8, 15768 (2017)".

[7] "M. Y. Gureev, A. K. Tagantsev, and N. Setter, "Head-to-head and tail-to-tail 180◦ domain walls in an isolated ferroelectric" Phys. Rev. B 83, 184104 , 2011".

[8] "P. S. Bednyakov, B. I. Sturman, T. Sluka, et al., "Physics and applications of charged domain walls", npj Comput Mater 4, 65 (2018)".

[9] "K. Moore, M. Conroy, E. N. O'Connell, et al., "Highly charged 180 degree head-to-head domain walls in lead titanate", Commun Phys 3, 231 (2020)".

[10] "R. K. Vasudevan et al., "Domain Wall Geometry Controls Conduction in Ferroelectrics", Nano Lett. 2012, 12, 11, 5524–5531".

[11] "J. Wang, et al, "Ferroelectric domain-wall logic units", Nat. Comm. volume 13, Article number: 3255 (2022)".

[12] "T. Böscke, J. Müller, D. Bräuhaus, U. Schröder, U. Böttger, "Ferroelectricity in hafnium oxide thin films", Appl. Phys. Lett. 99, 102903".

[13] "J. Müller et al., "Ferroelectric hafnium oxide: A CMOS-compatible and highly scalable approach to future ferroelectric memories," 2013 IEEE International Electron Devices Meeting, Washington, DC, USA, 2013, pp. 10.8.1-10.8.4".

[14] "J. Mueller, T. S. Boescke, U. Schroeder, S. Mueller, D. Braeuhaus, U. Boettger, L. Frey, T. Mikolajick, "Ferroelectricity in Simple Binary ZrO2 and HfO2", 2012 Nano Lett. 12 4318".

[15] "H.-J. Lee, M. Lee, K. Lee, J. Jo, H. Yang, Y. Kim, "Scale-free ferroelectricity induced by flat phonon bands in HfO2", Science, vol. 369, pp. 1343, 2020".

[16] "D. -H. Choe, S. Kim, T. Moon, S. Jo, H. Bae, "Unexpectedly low barrier of ferroelectric switching in HfO2 via topological domain walls", Materials Today, vol. 50, pp.8, 2021".

[17] "W. Ding, Y. Zhang, L. Tao, Q. Yang, Y. Zhou, "The atomic-scale domain wall structure and motion in HfO2-based ferroelectrics: A first-principle study", Acta Materialia, vol. 196, pp. 556, 2020".

[18] "T. Kiguchi, T. Shiraishi, T. Shimizu, H. Funakubo and T. J. Konno," Domain orientation relationship of orthorhombic and coexisting

monoclinic phases of YO1.5-doped HfO2 epitaxial thin films", Jpn. J. Appl. Phys. 57, 2018".

[19] "P. Zhou, B. Zeng, W. Yang, J. Liao, F. Meng, Q. Zhang, L. Gu, S. Zheng, M. Liao, Y. Zhou, "Intrinsic 90° charged domain wall and its effects on ferroelectric properties", Acta Materialia, Volume 232, 117, 2022".

[20] H. Tan, T. Song, N. Dix, F. Sánchez, I. Fina, "Vector piezoelectric response and ferroelectric domain formation in Hf0.5Zr0.5O2 films", Journal of Materials Chemistry C, vol. 11, pp. 7219-7226, (2023).

[21] "A. K. Saha, M. Si, K. Ni, S. Datta, P. D. Ye and S. K. Gupta, "Ferroelectric Thickness Dependent Domain Interactions in FEFETs for Memory and Logic: A Phase-field Model based Analysis," 2020 IEDM, San Francisco, CA, USA, 2020, pp. 4.3.1-4.3.4".

[22] "R. Koduru, A. K. Saha, M. Si, X. Lyu, P. D. Ye and S. K. Gupta, "Variation and Stochasticity in Polycrystalline HZO based MFIM: Grain-Growth Coupled 3D Phase Field Model based Analysis," 2021 IEDM, San Francisco, CA, USA, 2021, pp. 15.2.1-15.2.4".

[23] "R. Materlik, C. Künneth, A. Kersch; The origin of ferroelectricity in Hf1−xZrxO2: A computational investigation and a surface energy model. J. Appl. Phys. 7 April 2015; 117 (13): 134109".

[24] "T. K. Paul, A. K. Saha, S. K. Gupta, "Direction-Dependent Lateral Domain Walls in Ferroelectric Hafnium Zirconium Oxide and their Gradient Energy Coefficients: A First-Principles Study", Adv. Electron. Mater. 2023, 2300400".

[25] "K. D. Kim, et al., "Ferroelectricity in undoped-HfO2 thin films induced by deposition temperature control during atomic layer deposition", Journal of Materials Chemistry C, 2016,4, 6864-6872".

[26] "S. Oh, H. Kim, A. Kashir, H. Hwang, "Effect of dead layers on the ferroelectric property of ultrathin HfZrOx film", Appl. Phys. Lett. 117, 252906 (2020)".

[27] "Y. Goh, S. H. Cho, S.-H. K. Park, S. Jeon, "Oxygen vacancy control as a strategy to achieve highly reliable hafnia ferroelectrics using oxide electrode", Nanoscale, 2020,12, 9024-9031".

[28] "A. Acosta, J. Mark P. Martirez, N. Lim, J. P. Chang, E. A. Carter,"Relationship between ferroelectric polarization and stoichiometry of HfO2 surfaces", Phys. Rev. Materials 5, 124417".

[29] "M. Dogan, N. Gong, T. P. Maae, and S. I. Beigi,, "Causes of ferroelectricity in HfO2-based thin films: an ab initio perspective", Physical Chemistry Chemical Physics, vol. 21, pp. 12150, 2019".

[30] "J. Sifuna, et al., "First-principles study of two-dimensional electron and hole gases at the head-to-head and tail-to-tail 180 domain walls in PbTiO3 ferroelectric thin films", Phys. Rev. B, 101, 174114, 2020".

[31] "S. V. Kalinin, Y. Kim, D. D. Fong, A. N. Morozovska, "Surface-screening mechanisms in ferroelectric thin films and their effect on polarization dynamics and domain structures", Rep. Prog. Phys. 2018, 81 036502".

[32] "S. Hong, S. M. Nakhmanson, D. D Fong, "Screening mechanisms at polar oxide heterointerfaces", Rep. Prog. Phys. 2016, 79 076501".

[33] "Y. Choi, H. Park, C. Han, J. Min, C. Shin, "Improved remnant polarization of Zr-doped HfO2 ferroelectric film by CF4/O2 plasma passivation", Sci Rep. 2022 Oct 6;12(1):16750".

[34] "R. R. Mehta, B. D. Silverman, J. T. Jacobs; Depolarization fields in thin ferroelectric films. J. Appl. Phys. 1 August 1973; 44 (8): 3379–3385".

[35] "D. J. Kim, J. Y. Jo, Y. S. Kim, Y. J. Chang, J. S. Lee, Jong-Gul Yoon, T. K. Song, T. W. Noh, "Polarization Relaxation Induced by a Depolarization Field in Ultrathin Ferroelectric BaTiO3 Capacitors", Phys. Rev. Lett. 2005, 95, 237602".

[36] "P. Giannozzi, et al., "QUANTUM ESPRESSO: a modular and open-source software project for quantum simulations of materials", J.Phys.: Condens.Matter 21, 395502 (2009)".

[37] "A. Kokalj, "Computer graphics and graphical user interfaces as tools in simulations of matter at the atomic scale", Comp. Mater. Sci., 2003, 28, 155-168".

[38] "N. A. Spaldin, "A beginner's guide to the modern theory of polarization", J. of Solid State Chem. 2012, 195, 2.".

[39] "M. H. Park, et al. ,"Effect of Zr Content on the Wake-Up Effect in Hf1–xZrxO2 Films", ACS

Applied Materials & Interfaces 2016 8 (24), 15466-15475".

[40] "J. Lyu, I. Fina, R. Solanas, J. Fontcuberta, F. Sánchez; "Robust ferroelectricity in epitaxial Hf1/2Zr1/2O2 thin films", Appl. Phys. Lett. 20 August 2018; 113 (8): 082902".

# Supplementary Information

# Head-to-Head and Tail-to-Tail Domain Wall in Hafnium Zirconium Oxide: A First Principles Analysis of Domain Wall Formation and Energetics

Tanmoy K. Paul, Atanu K. Saha, Sumeet K. Gupta
*Purdue University, West Lafayette, Indiana, 47907, USA*
*Email: paul115@purdue.edu / Phone: (765) 607-3147*

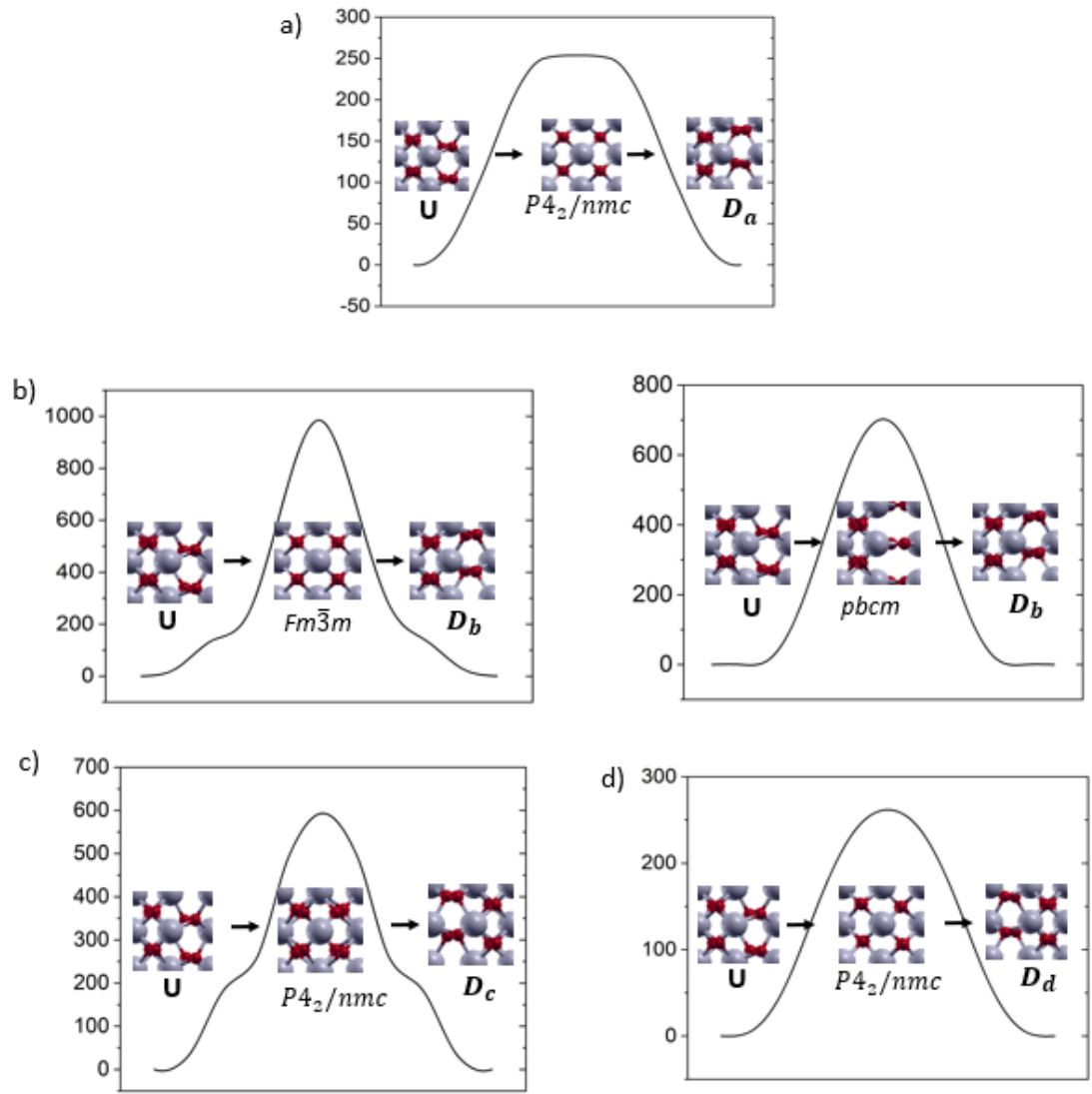

**Figure S1** a)-d) Polarization reversal path for upward polarized unit cell, U to downward polarized unit cells $D_a$, $D_b$, $D_c$ and $D_d$ respectively. The unit cells are taken from Figure 1 of main text. From U to $D_b$ transition, both $Fm\bar{3}m$ and *pbcm* can occur as intermediate state as shown in the two paths of b). For U to $D_a$, $D_c$ and $D_d$ transition, $P4_2/nmc$ occurs as the intermediate phase.

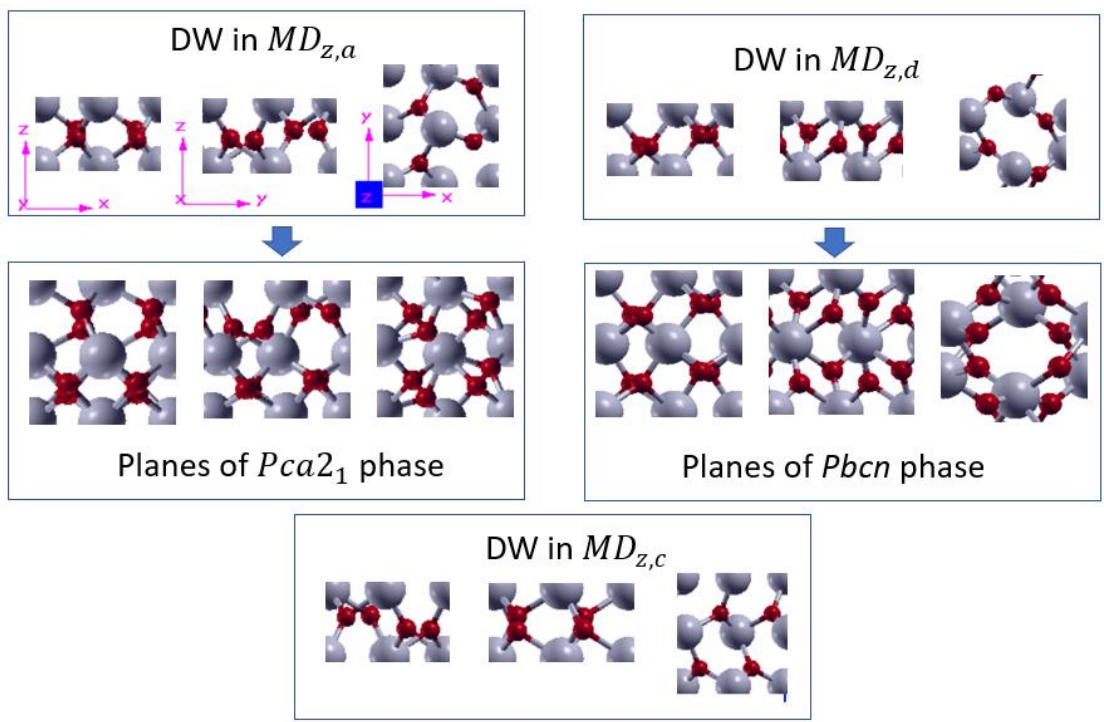

**Figure S2** Atomic configuration of DWs in Figure 2. All the DWs are half unit cell thick and so half-cell thick configuration is shown here. The DWs in $MD_{z,a}$ and $MD_{z,c}$ are orthorhombic $Pca2_1$ phase and 90° rotated with respect to the domains. The DW in $MD_{z,d}$ resemble orthorhombic *pbcn* phase. For better comparison purpose, planes of full unit cell of $Pca2_1$ and *pbcn* phases are also demonstrated.

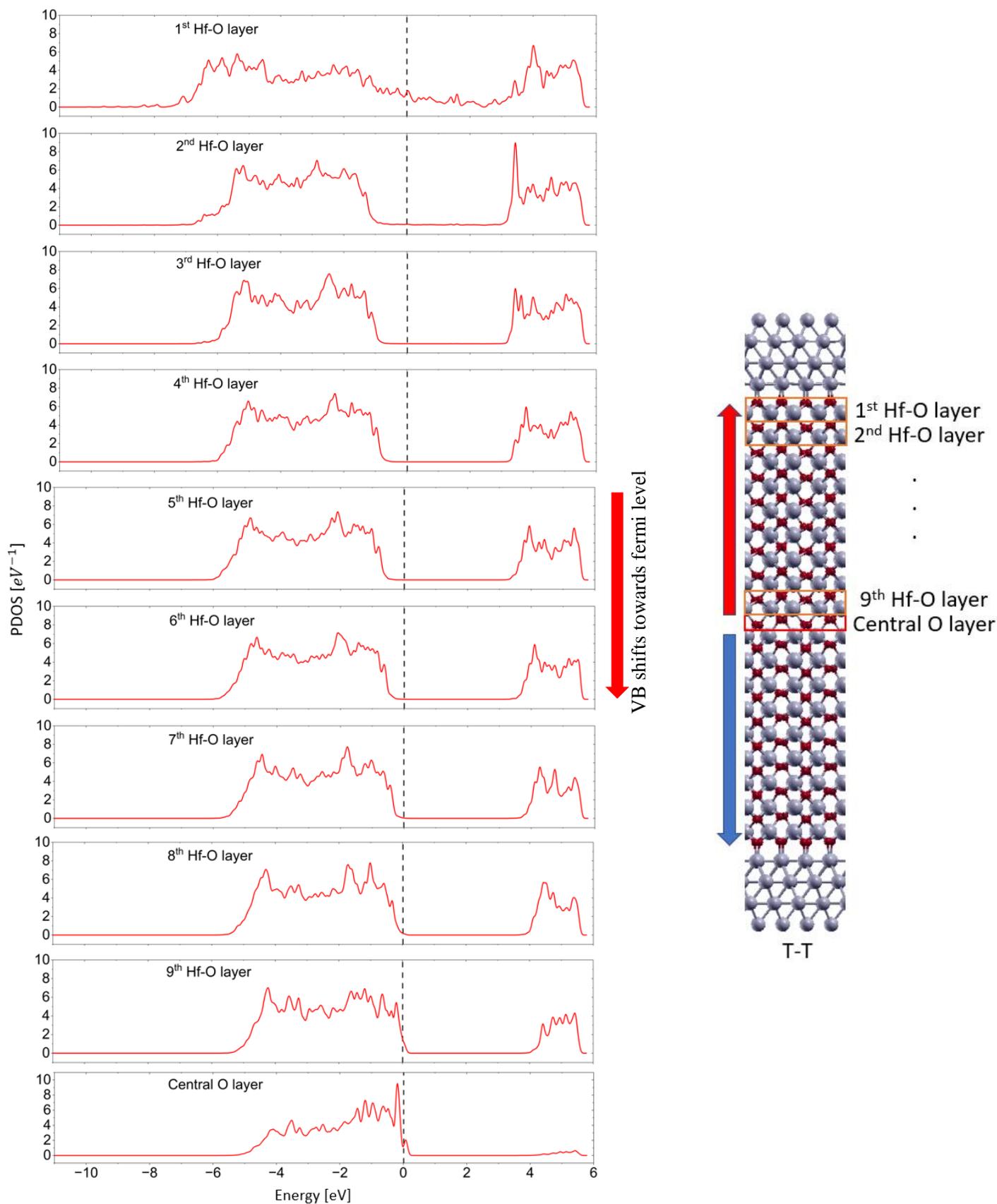

**Figure S3** Layer wise PDOS from the interfacial Hf-O atoms towards the central O atoms at the DW for T-T configuration. VB shifts towards the fermi level as we move towards the DW region which indicates that holes appear near the DW. (The first Hf-O layer is interfacial; therefore, they are metalized and corresponding projected bandgap is not observable.)

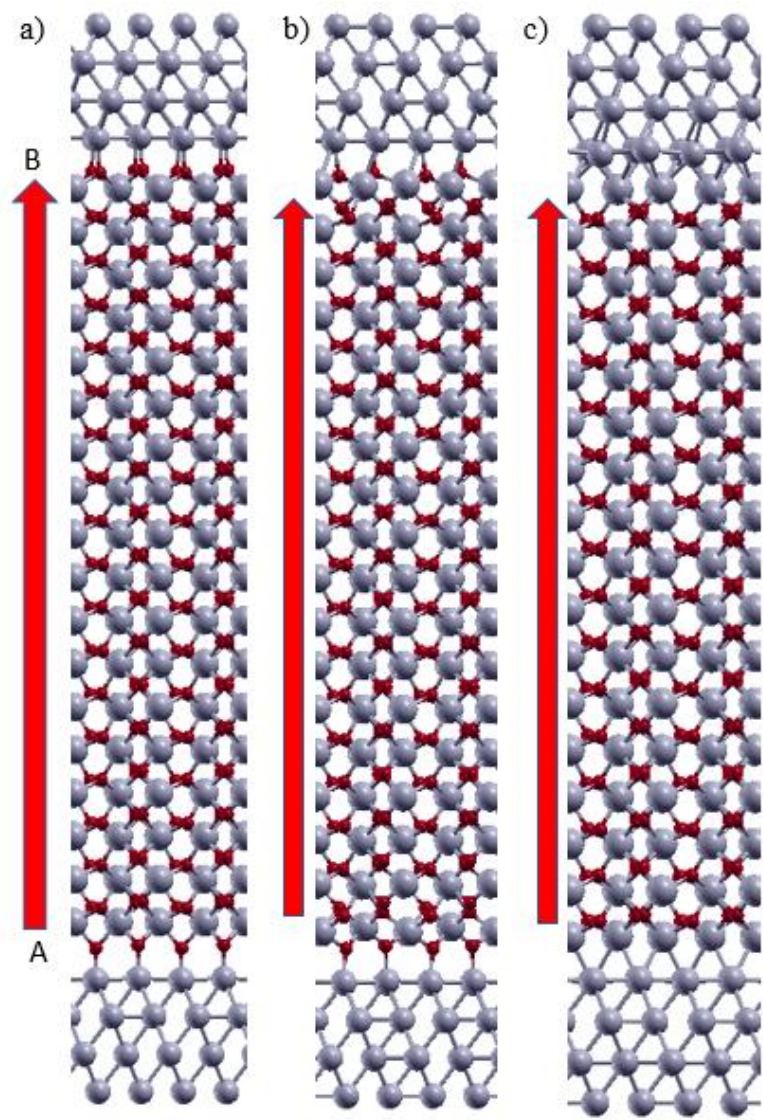

**Figure S4** Atomic configuration of single domain structures with a) vacancy-free O-ended interfaces b) interfaces with one spacer and one polar vacancy c) Hf/Zr-ended interfaces. A and B interfaces are marked as bottom and top interfaces respectively.

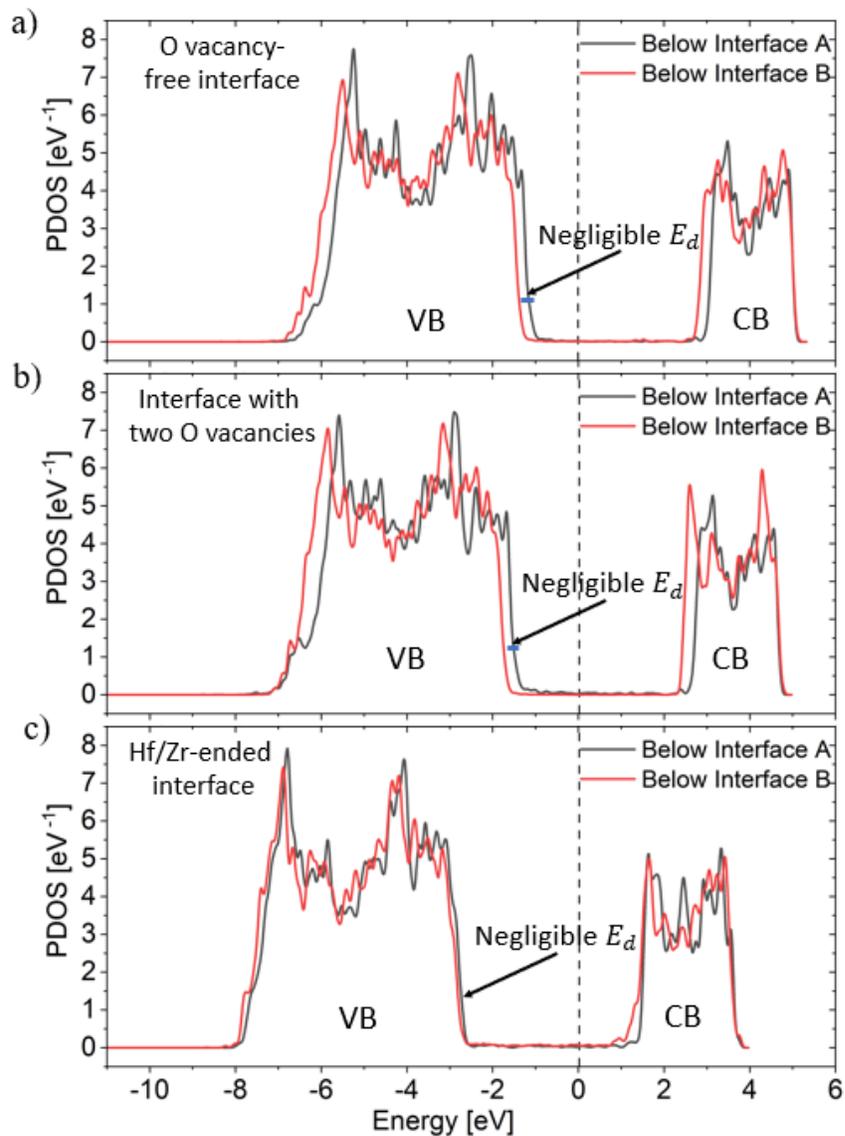

**Figure S5** Projected Density of States (PDOS) of atomic layers of SD HZO adjacent to the two interfaces for a) O-ended interfaces without vacancy b) interfaces with one spacer and one polar vacancy c) Hf/Zr-ended interfaces shown in Figure S4a-c respectively. The negligible shift between the bands near the two interfaces dictates negligible depolarization field, i.e., excellent bound charge screening by metal irrespective of oxygen vacancy density.

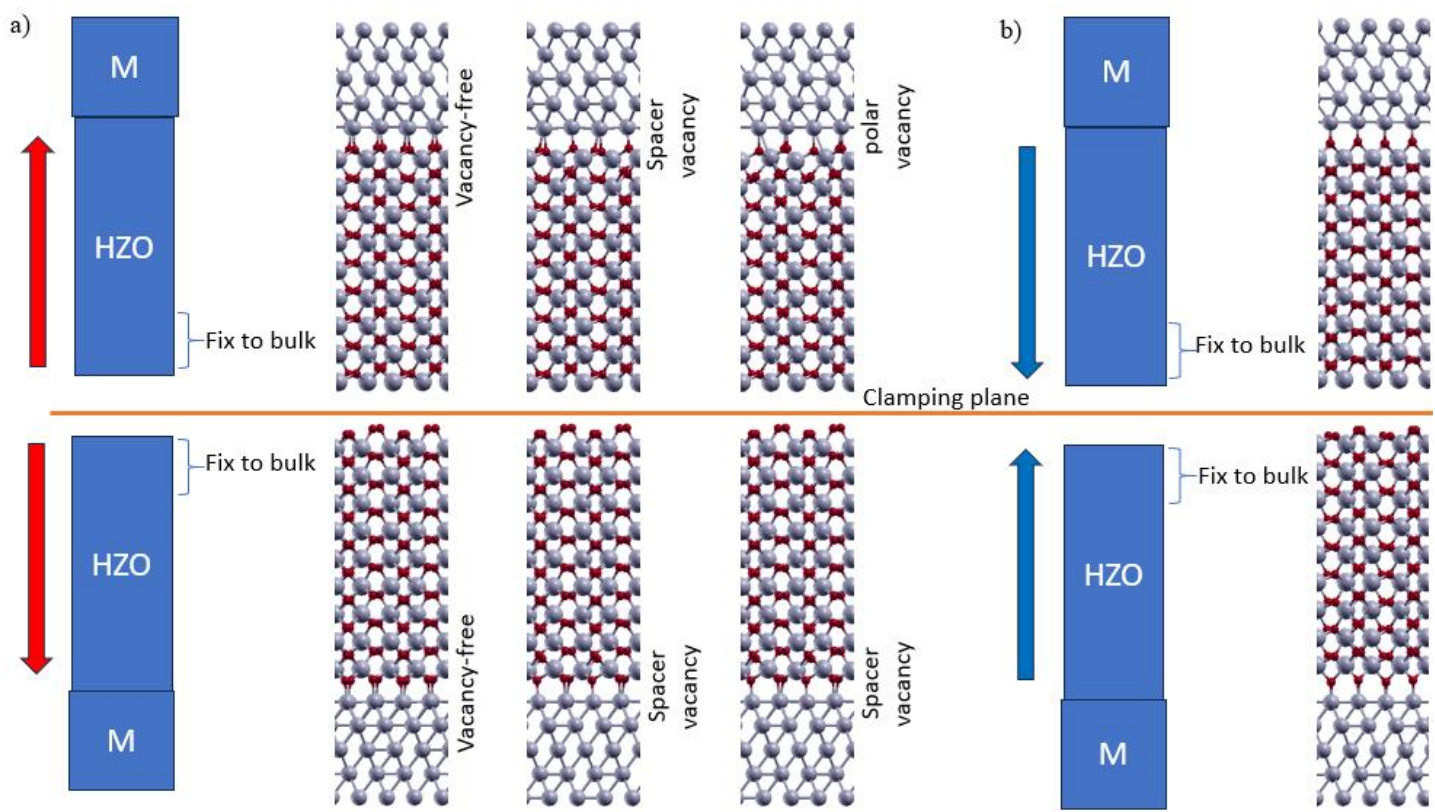

**Figure S6** a) Metal-HZO stacks to form T-T configuration for vacancy-free O-ended interface and interfaces with single oxygen vacancy in either polar or spacer layer. b) Metal-HZO stacks to form H-H configuration for vacancy-free O-ended interface. To obtain relaxed interfaces of each stack, the furthest HZO unit cell from each of the interfaces are kept fixed to bulk atomic configurations. The in-plane lattice parameters are kept fixed to HZO and out-of-plane lattice parameter is relaxed with the presence of a vacuum on both sides. The top and bottom stacks are clamped together in the orange plane to get initial (not fully relaxed) structures of T-T and H-H configurations.

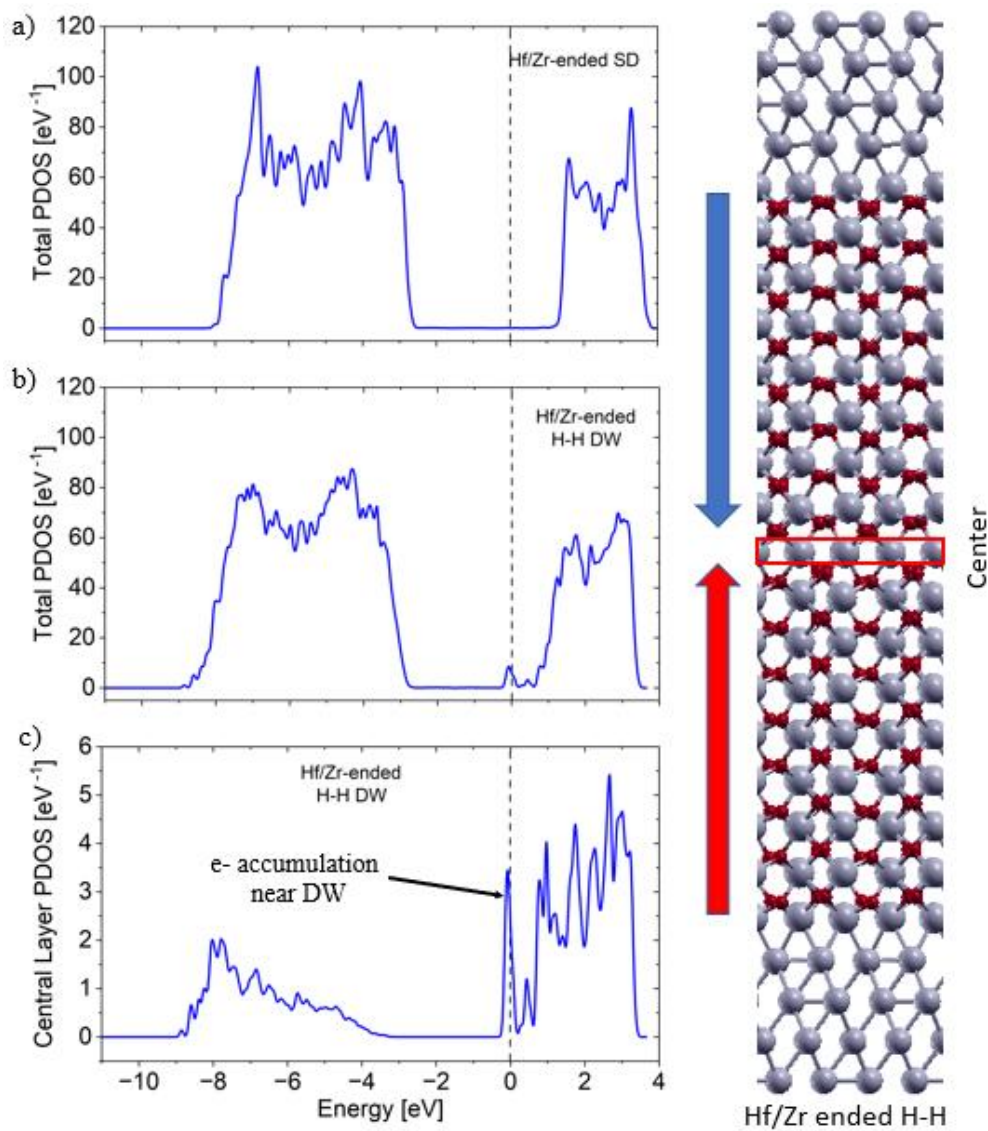

**Figure S7** Total PDOS of atomic layers of a) SD HZO and b) MD HZO of H-H configuration for Hf/Zr-ended interfaces c) PDOS at central Hf/Zr layer for H-H DW configuration in HZO. The CB minimum and fermi level suggest that electrons appear at the DW to screen positive polarization bound charges. This comes at the cost of depolarization field inside the domains. The structure at the right side shows the abovementioned H-H structure.